\documentclass[aps,prb,twocolumn,groupedaddress,amsmath,amssymb]{revtex4}
\usepackage{graphicx}
\usepackage{calc}
\usepackage{psfrag}
\usepackage{dcolumn}
\newcommand{\abst}[1]{{\left| #1 \right|}^2}
\newcommand{\evec}[1]{\underline{\widehat{#1}}}
\newcommand{\Gfak}{{\frac{e^2}{\pi \hbar}}}
\newcommand{\kapta}{{\kappa_{\vta}}}

\newcommand{\kb}{{k_\mathrm{B}}}
\newcommand{\kpar}{{\Vec{k}_{\|}}}
\newcommand{\kta}{{\Vec{k}_{\vta}^\alpha}}
\newcommand{\ktp}{{\Vec{k}_{\vta}^+}}
\newcommand{\ktpp}{{\Vec{k}_{\vtap}^+}}
\newcommand{\Mat}[1]{{\mathbf{#1}}}
\newcommand{\mathp}{{\; .}}
\newcommand{\mathk}{{\; ,}}
\newcommand{\mOm}{{$\mu\Omega$cm}}
\newcommand{\ocite}[1]{{Ref.~\onlinecite{#1}}}
\newcommand{\PLi}[1]{{{}^L \Psi_{#1}^{inc}}}
\newcommand{\PLs}[1]{{{}^L \Psi_{#1}^{sc}}}
\newcommand{\Ppi}[1]{{{}^p \Psi_{#1}^{inc}}}
\newcommand{\Pps}[1]{{{}^p \Psi_{#1}^{sc}}}
\newcommand{\R}{{\Mat{R}}}
\newcommand{\Res}{{\mathcal{R}}}
\newcommand{\Resfak}{{\frac{\pi \hbar}{e^2}}}
\newcommand{\T}{{\Mat{T}}}
\newcommand{\lra}{\left\langle}
\newcommand{\rra}{\right\rangle}
\newcommand{\refeq}[1]{Eq.~(\ref{#1})}
\newcommand{\reffig}[1]{Fig.~\ref{#1}}
\newcommand{\refk}[1]{(\ref{#1})}
\newcommand{\reftab}[1]{Tab.~\ref{#1}}

\newcommand{\vta}{{\Vec{\tau}}}
\newcommand{\vtap}{{\Vec{\tau}'}}
\newlength{\picincolumnwidth}
\psfrag{DOSLabel}{\hspace*{-5ex}tot. DOS /[states/eV]}
\psfrag{ELabelx}{$(E - E_\mathrm{F})/$[Ry]}
\psfrag{ELabelx1}{$(E - E_\mathrm{F})/$[eV]}
\psfrag{GLabely}{$G/ \Gfak$}
\psfrag{kpaL}{$\ktp$}
\psfrag{kpaR}{$\ktpp$}
\psfrag{xLabelL}{$L/$[au]}
\psfrag{xLabelRad}{$r/$[au]}
\psfrag{yLabelPaar}[l]{$g(r)$}
\psfrag{yLabelRes}{$a^2 \Res/[\Resfak \mathrm{au}^2]$}
\psfrag{yLabelRes1}{$\Res,\, \lra G \rra^{-1}/[\Resfak]$}
\psfrag{yLabelVarG}{var$(G/\Gfak)$}
\psfrag{labelrhok}{$\rho/$[\mOm]}
\psfrag{labelkpx}[tl][][1][-5]{$k_{\|x}/ [\pi/a]$}
\psfrag{labelkpy}[br][][1][35]{$k_{\|y}/ [\pi/a]$}
\begin{document}


\title{Calculation of resistance for weak scattering, strong scattering and insulating 
quasi-one dimensional systems}
%

\author{Andr\'e L\"oser}
\email[]{loeser@physik.tu-chemnitz.de}
\affiliation{Technical University Chemnitz, Institute of Physics, D-09107 Chemnitz, Germany}


\date{\today}

\begin{abstract}
A parameter free calculation of the resistivity is applied to liquid metals near the melting point
ranging from weak to strong scattering limit.
The method is based on length dependent resistance calculations for quasi-one dimensional systems
and was applied on structures with up to 10000 atoms.
The calculated value for conductance fluctuations is in good  agreement with theoretical predictions.
The resistivities are 
compared with the Kubo-Greenwood and the extended Ziman formula with the same scattering
potential and similar structure.
The resistance calculation is applicable for insulating materials as well, 
which is demonstrated for crystalline and amorphous silicon.
\end{abstract}


\maketitle
%
%
%
\section{Introduction}
Electronic transport in mesoscopic systems is marked by interference effects, which can not be
understood by classical considerations.
Effects such as universal conductance fluctuations (UCF), Aharonov-Bohm effect, resonant scattering, 
etc.
have been investigated for years.
One of the most striking effects is the localization in pure one dimensional
disordered systems.
This led to the scaling approach,\cite{Abrahams79}
which demonstrated disorder induced transition from metallic to insulating behavior 
(Anderson transition)
for three dimensional systems.

Starting from one dimensional systems\cite{Kumar85} 
a scaling approach was developed for quasi-one dimensional
systems employing random scattering matrices.\cite{Beenakker97}
The length-dependence of the scattering system is described by the diffusion-type 
DMPK\cite{Dorokhov82,Mello88} equation, which has a metallic solution for lengths shorter then the
localization length.
The resistance increases linear with the sample lengths apart from the fluctuations.
For large lengths the system behaves like a one-dimensional system with a localization length
dependent on sample size (number of channels) and the elastic mean free path.

The resistance calculations are based on formulas introduced for one-dimensional systems
by Landauer\cite{Landauer57} and extended later to 
quasi-one dimensional systems\cite{Fisher81,Buttiker85}.
In the simplest case the conductance is proportional to the total current 
transmission through the sample,
$G \sim T$, which can be understood as ``measurement'' of the current for a defined
voltage drop over the sample.
A somewhat different approach 
was followed by Lenk\cite{Lenk94} using a functional of the currents.

For practical applications
the nearly classical scaling of the resistance was used by Kahnt\cite{Kahnt95} to calculate the
resistivity for quasi-one dimensional systems of muffin-tin scatterer.
Subsequently a similar approach was used to calculate resistivities
within the tight-binding formalism.\cite{Todorov96}
Both approaches were hampered by large deviations from the classical scaling 
due to the limited system size.

This paper follows the approach of \ocite{Kahnt95}, because it is a parameter-free description
of real materials.
Compared to previous work the calculation of the scattering matrix is done in a combination of plane
wave and angular momentum representation.
This hybrid representation preserves the high precision of the angular moment representation, 
which is
necessary for resistance calculations, but on the other hand avoids the cubic increase in the
computation time. The computation time increases linear with the sample length.
Thus up to 10000 atoms can be computed with angular momentum channels up to $l=2$. 
This makes it
possible to extend the resistivity calculation on weak scattering systems. 

In section \ref{resist} the calculation of the scattering matrix and the resistance is outlined.
The details of the hybrid method are given in appendix~\ref{app_hyb}.

The sample length dependence of the resistance is analyzed for metallic samples
in section \ref{metall}.
It is shown, that there are three different regimes for the length-dependent resistance.
The theoretical value for the UCF is reproduced.
An upper bound for the conductance is found for the localized regime.
The section is concluded with practical aspects of the resistivity calculation, e.g. temperature
dependence.

The method of resistivity calculation is applied to liquid metals near the melting point
in section \ref{example}.
Calculations for strong scattering 3d transition metals and weak scattering liquid sodium 
are presented.

Usually the resistivity would be calculated by Kubo-Greenwood formula\cite{Kubo57,Greenwood58} or 
in case of weak scattering 
by Ziman formula\cite{Ziman61}, therefore the presented method is compared to these methods using 
similar input
data, e.g. structure and scattering potentials.
In both cases there is a good agreement. Small but systematically larger values are found for the
transition metals, which may be connected to weak scattering corrections.

For weak scattering much larger samples are necessary. 
It is shown, that the numerical effort can be
reduced by using an alternative resistance ``measurement'' employing adaptive leads.
The resistivity as well as the resistivity coefficient are calculated 
near the melting point and compared to measurements and
calculation with  the extended Ziman formula\cite{Evans71}
using the same structure factor and scattering potentials.
The influence of multiple scattering is discussed.

The method is finally applied to crystalline and amorphous silicon. 
These insulating samples 
show a different length dependence of the resistance, where
in contrast to localized quasi-one dimensional metallic samples,
the localization length is
characteristic for the insulating material.
%
%
%
%
%
%
\section{\label{resist}Calculation of resistance and resistivity}
\subsection{\label{scat}The scattering matrix of the stack}
The scattering matrix is needed for the calculation of the resistance.
For most applications the sample length dependence of the resistance has to be known. 
Therefore successive calculation of the scattering matrix is employed to minimize the computation 
time.
Usually the computation is done atom layer by atom layer for a stack of layers up to 10000
atoms per sample.

The scattering of the atoms is approximated by a muffin-tin potential calculated self-consistent by a
LMTO method in local density approximation (LDA).\cite{Arnold95}
The scatterers are described by their energy dependent phases shift $\eta_l (E)$ for the
relevant angular momentum $l$.
This energy dependence is droped in the following equations, because
the multiple scattering is calculated for one energy $E = k^2$.

Periodic boundary conditions are used in lateral direction (perpendicular to the current). 
Therefore the wave field outside is best represented in plane waves
\begin{equation}
\label{scat_eq:1}
\varphi_{\vta}^{\alpha}(E,\kpar,\Vec{r}) =
\frac{1}{\sqrt{\kappa_\vta}}\exp{i\Vec{k}_\vta^{\alpha} \Vec{r}}
\end{equation}
with wave vector
\begin{equation}
\Vec{k}_\vta^{\alpha} = \kpar + \vta + \evec{e}_z \alpha \kapta
\end{equation}
and the component in direction of the flowing current
\begin{equation}
\kapta=\sqrt{k^2-(\kpar+\vta)^2}
\mathk
\end{equation}
where $\alpha=\pm$ denotes right or left moving waves.
Note the dependence on the parallel wave vector $\kpar$, due to the periodic boundary conditions in
lateral direction.
This vector can assume values in a two dimensional Brillouin zone due to the lateral boundary
conditions.
The plane wave itself is identified by the two dimensional vector $\vta$.


There are propagating waves and evanescent waves with real and imaginary impulse 
$\kapta$
perpendicular to the stack. 
Whereas the number
of propagating waves $N \approx \frac{k^2}{4\pi} a^2$ is finite, 
the number of evanescent waves is not limited.
But only a finite number of the
evanescent waves contribute to the total current,
because of the exponential decreasing amplitude of these waves. 
This number increases with decreasing distance
between neighboring layers of atoms.\cite{Kahnt95}

The scattering of the stacked layers is described by the scattering matrix $\Mat{S}$.
The left and right outgoing waves are given by the left and right incident plane waves
\begin{equation}
\label{scat_eq:2}
\left|
\begin{array}{c}
\Vec{\Psi}^+_r \\
\Vec{\Psi}^-_l \\
\end{array}
\rra
= 
\left[ 
	\begin{array}{cc}
		\Mat{S}^{++} & \Mat{S}^{+-}  \\
		\Mat{S}^{-+} & \Mat{S}^{--}  \\
	\end{array} 
\right]
\left|
\begin{array}{c}
\Vec{\Psi}^+_l \\
\Vec{\Psi}^-_r \\
\end{array}
\rra
\mathk
\end{equation} 
where for example $\Vec{\Psi}^\pm_l$ denotes the amplitudes of the incident and outgoing 
plane waves on the left of the stack.
Only the propagating part of the matrix $\Mat{S}$ is used for the resistance calculation.
These matrix elements are the transmission and reflection amplitudes of the stack.
To distinguish with the scattering matrix in \refeq{scat_eq:2}, the transmission and reflection
amplitudes will be denoted by $\Mat{t}$ and $\Mat{r}$ respectively.
Note that only this propagating part of the scattering matrix is unitary.

\begin{figure}
\includegraphics[width=\picincolumnwidth]{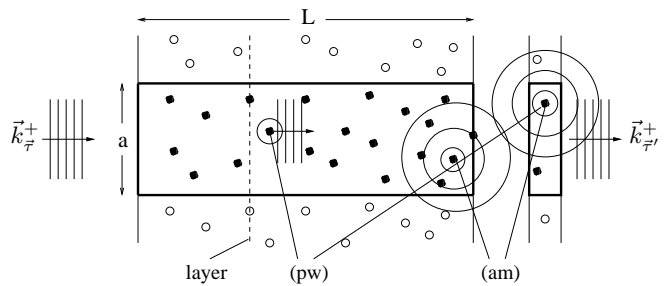}
\caption{\label{scat_pic:1}%
Two dimensional outline of the simultaneous use of plane waves (pw) and angular momentum (am)
representation of the wave field.
The stack of layers with lateral length $a$ and stack length $L$ is increased by an additional
sub-stack.}
\end{figure}
The successive calculation of the scattering matrix can be done easily in this plane wave 
representation by
eliminating the wave field between the main stack and the additional layers 
(\reffig{scat_pic:1}).\cite{Cahay88}
The computation time increases only linear with
the number of layers. 
This is the method of choice, when the number of necessary plane waves is small.
The plane wave representation is not applicable
on samples of highly disordered material with large lateral length $a$ and high angular
momentum channels for the scatterers.

To circumvent this problem the successive calculation of the scattered waves was
done in angular momentum representation in \ocite{Kahnt95}.
The propagation between the atoms is described by structure constants $G^{ij}_{LL'}$ 
given in appendix~\ref{app_prop}.
The scattering is defined by the scattering amplitudes $2 i k f_{iL} = 1 - e^{2i\eta_l}$, where
$L=(l,m)$ denotes the angular momentum channel at the $i$-th atom.
The wave field in angular momentum basis is given by
\begin{equation}
\label{scat_eq:4}
\left| \Psi \rra = \left| \Psi^{inc} \rra +
\Mat{G} \Mat{F} \left| \Psi^{} \rra
\mathk
\end{equation}
where a compact matrix notation is used.
The wave field is calculated by a successive matrix inversion.  

The plane wave representation for the stack is reached by the projection from plane wave to angular
momentum channels for the incident waves and vice versa for the scattered waves,\cite{Kahnt95} 
\begin{equation}
\label{scat_eq:3}
\Mat{S} = P^{\rm hom} + i k \Mat{B}
\left[ \Mat{F}^{-1} - \Mat{G} \right]^{-1} \Mat{A}
\mathp
\end{equation}
The draw back of the angular momentum representation 
is an increasing number of scattering channels with increasing numbers of
layers. This limits the number of atoms to a few hundred for d-scattering. 

As will be shown below, for the calculation of material specific properties, e.g. the resistivity,
larger samples with more then 1000 layers are necessary to achieve a good accuracy compared to 
Kubo-Greenwood calculations.

Therefore a combination of plane wave and angular momentum basis was 
used to represent the wave field in the stack.\cite{MacLaren893}
Whereas far away parts of the stack are projected on plane waves, the scattering waves of 
close layers are dealt with in angular momentum basis.
This limits both the number of plane waves and angular momentum channels.
The computation time increases only linear with the number of layers.
The details are given in the appendix~\ref{app_hyb}.
\subsection{\label{res}Calculation of the resistance}
The resistance is usually calculated by a multi-channel B\"uttiker formula\cite{Fisher81}
\begin{equation}
\label{res_eq:1}
G = \frac{e^2}{\pi \hbar}  T
\mathp
\end{equation}
which connects the conductance $G$ with the total current transmission through the stack of
layers.\footnote{The factor 2 for the electron spin is already included.}

The total transmission $T$ is calculated from the transmission amplitudes $t_{ij}$ 
of \refeq{scat_eq:2}.
Because no scattering takes place in the semi-infinite leads to the reservoirs, the current flows only
through the $N$ propagating channels, 
\begin{equation}
T = \sum_{i=1}^N \sum_{j=1}^N \abst{t_{ij}}
\mathp
\end{equation}
A shortcoming of \refeq{res_eq:1} is the non-vanishing resistance $\Res_0 = \Resfak N^{-1}$,
if there are no scattering layers at all.
The resistance formula \refeq{res_eq:1} is not unique, due to the boundary conditions left and right
of the stack.
The ideal leads in connection with the reservoirs define an even 
incident current distribution.\cite{Lenk90}

To evaluate the influence of the left and right boundary conditions on the resistance and more
important on the resistivity, we compare the ideal leads in case of \refeq{res_eq:1} with an ansatz
employing adaptive leads,\cite{Lenk94}
\begin{equation}
\label{res_eq:2}
G = \, \frac{e^2}{\pi \hbar}
\sum_{i=1}^N \sum_{j=1}^N \tilde{T}_{ij}
\mathk
\end{equation}
using the elements of the matrix 
\begin{equation}
\label{res_eq:2b}
\Mat{\tilde{T}} = 2\; \Mat{T} \left[ \Mat{1} + \Mat{R} - \Mat{T} \right]^{-1}
\mathp
\end{equation}
The matrix elements for the transmission $T_{ij} = \abst{t_{ij}}$ and 
reflection probabilities $R_{ij} = \abst{r_{ij}}$ are given by the transmission
and reflection amplitudes respectively.
\begin{figure}[t]
\includegraphics[width=\picincolumnwidth]{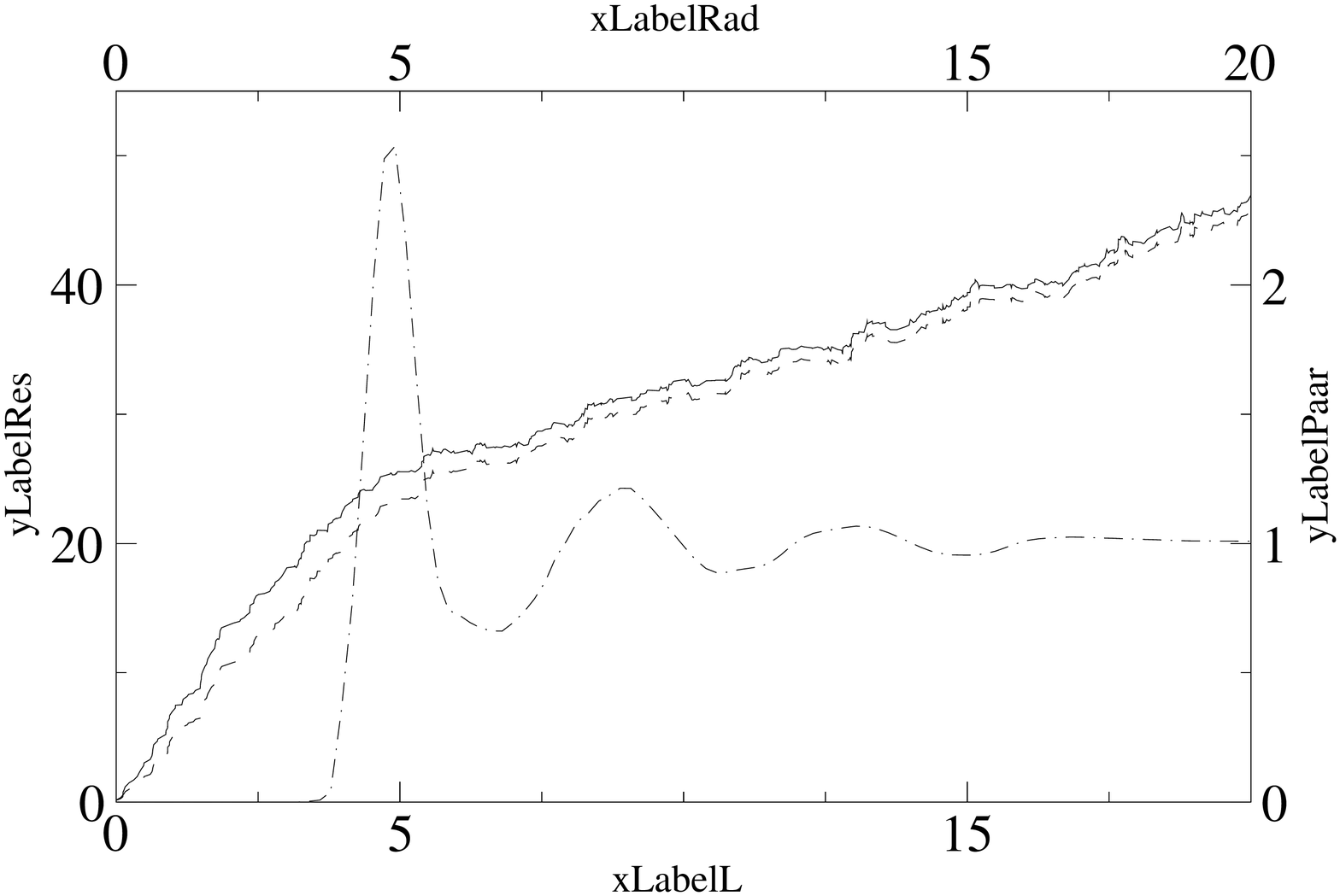}
\caption{\label{length_pic:1}%
Comparison of the length-dependent resistance with the radial pair correlation function $g(r)$
(-- $\cdot$ --) shows the
influence of the nearest neighbor distance for strong scattering. 
The resistance was calculated with \refeq{res_eq:1} (---) and \refk{res_eq:2} (-- --) in
$\Gamma$-point approximation at the Fermi energy.
The residual resistance $\Res_0$ is omitted for \refeq{res_eq:1}.
The lateral length of the sample was $a=40$ au.
Liquid iron near 
the melting point was taken as a typical example for all 3d transition metals. 
The radial pair correlation was taken from \ocite{Waseda80}.}
\end{figure}

\refeq{res_eq:2} gives a minimal resistance with respect to a given functional of the incident
currents.\cite{Lenk94}
In that sense it is a good test for the deviations of \refeq{res_eq:1}.
The inverse matrix in \refeq{res_eq:2b} characterizes the adaption of the currents to the scattering
by the stack.

From \reffig{length_pic:1} it can be seen, that differences in the resistance occurred for short stack 
length.
But these differences in the resistance remain constant for larger stacks.
The length dependence is then dominated by the transmission.
Nevertheless the resistance in \reffig{length_pic:1} 
is always  smaller for adaptive leads then for ideal leads.
%
%
%
%
\section{\label{metall}Resistance for disordered metallic samples}
\subsection{\label{length}Dependence on the sample length}
For a disordered metallic sample the length dependence of the resistance can be divided in three
different regions.

For a short length the stack is not material specific.
This can be seen from the difference of the measured resistance, i.e. 
\refeq{res_eq:1} and \refk{res_eq:2}.
This not material specific region is given by the characteristic length scales of the material,
e.g. the correlation length and the elastic mean free path.

From \reffig{length_pic:1} most strikingly visible is the connection to the interatomic distance.
This is caused by the strong scattering between nearest neighbors in 3d transition metals.
This strong scattering enables an effective current transmission through the sample, 
if atoms are stacked one behind the other. 
Therefore the slope of the resistance is usually smaller beyond this initial region.

The second region is material specific. The resistivity can be calculated from the slope of the
resistance over the sample length. This is done by linear regression and can be seen as a fit to
the classical relation 
\begin{equation}
\label{length_eq:1}
\Res = \Res_b + \rho \frac{L}{A}
\end{equation}
for a metallic wire of length $L$ and cross section $A=a^2$.
The initial region discussed above is accounted for by a length independent resistance
$\Res_b$.

As can be seen from \reffig{length_pic:1}, there is no notable change in the difference between the
resistance calculated with \refeq{res_eq:1} and \refeq{res_eq:2} in the metallic region.
Therefore the linear regression with \refeq{length_eq:1} gives a ``measurement'' independent
resistivity.
Length dependent deviations for different ``measurements'' of the resistance indicate a non-material
specific behavior (see sec. \ref{weak}).

\begin{figure}
\includegraphics[width=\picincolumnwidth]{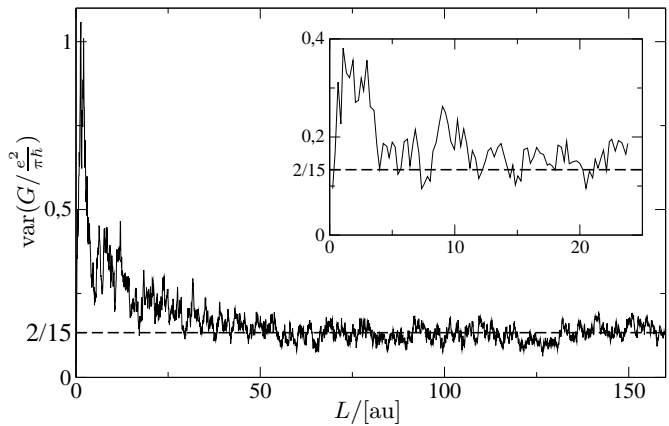}
\caption{\label{length_pic:2}%
Variance of the conductance for $a=40$ au and $a=20$ au (inlet) plotted over the sample length $L$.
The theoretical value is $2/15$.\cite{Beenakker97}
50 samples as shown in \reffig{length_pic:1} were used for $a=40$ au and 
all the samples for the inlet are shown in
\reffig{length_pic:4}.}
\end{figure}
The deviation from the classical behavior of \refeq{length_eq:1} is of order $\Gfak$ for the
conductance and 
known as Universal Conductance
Fluctuations (UCF). 
In \reffig{length_pic:2} the variance of the conductance is plotted over the sample length $L$
for two different lateral length $a$.
The smaller samples were calculated by the original method used in \ocite{Kahnt95}.
The hybrid representation of appendix~\ref{app_hyb} has to be used for the doubled lateral length.

For sample length larger then the lateral length, i.e. $L>a$, 
there is a good agreement with the theoretical value $2/15$ for UCF in quasi-one dimensional 
systems.\cite{Beenakker97,Lee87}
This agreement shows that the variations in the resistance curves stem from the coherent scattering
and not from variations of the underlying material.
This is important for calculations of material specific resistivities.

From the practical point of view,
the conductance fluctuations result in variations of the resistance slope, most 
notably for small later length. The resistivities for the smaller samples of \reffig{length_pic:2} 
lie between $65$ and $145$ \mOm, which makes a calculation of a single sample to unreliable for
comparison with experiment and Kubo-Greenwood method.

By doubling the lateral length the variation of the resistivity is
reduced below $\pm 10$ \mOm. 
Thus the material is much more defined and the resistivity can be calculated
from one sample.
An averaging over many samples is not necessary.

\begin{figure}[b]
\includegraphics[width=\picincolumnwidth]{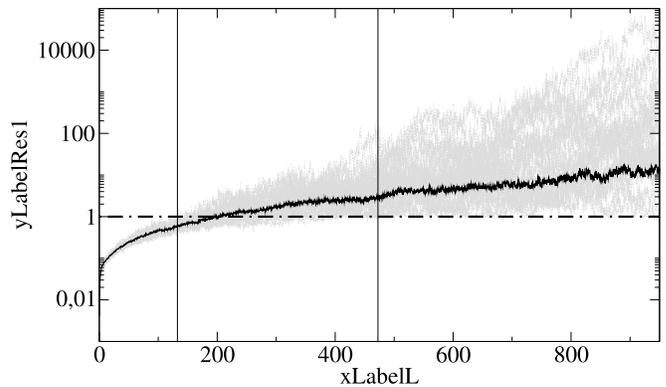}
\caption{\label{length_pic:3}%
The inverse average of the conductance $\lra G \rra^{-1}$ calculated with \refeq{res_eq:1} is plotted 
for long sample length $L$.
The resistance of all 24 samples are plotted with gray lines.
The boundary between metallic and localized region ($G = \Gfak$) is marked by (-- $\cdot$ --). 
The transition region, when the first and the last sample falls below this threshold ($G = \Gfak$), 
is outlined by vertical lines (comp. \reffig{length_pic:4}).
The liquid iron samples had a lateral length of $a\approx20$ au ($N=20$).}
\end{figure}
Anyway, if the length $L$ increases the conductance becomes of the same order as the fluctuations. 
In \reffig{length_pic:3} this is shown for many samples with small lateral length.

The average of the conductance decreases exponentially.
Different averaging schemes have been proposed and tested in \ocite{Todorov96}.
The quasi-one dimensional samples become localized due to the disorder just 
as disordered one dimensional systems for large $L$ .

From \reffig{length_pic:3} it can be seen, that $G = \Gfak$ is a lower boundary for the localized
samples.
If the total transmission of a specific sample falls below $T = 1$, 
then it does not increase beyond that limit for any larger length.
The wave field is characterized by the localization length 
$l_\mathrm{loc} \approx N l_\mathrm{el}$,\cite{Beenakker97}
where $l_\mathrm{el}$ is the elastic mean free path.
The resistance is not solely depending on the material like in the second regime,
but also on the lateral constrains and is therefore not material-specific.

In this localized regime the current is dominated by transmission through isolated 
quasi-bound states in contrast to the metallic regime, where many states contribute to the current.
If the resonance energy coincides with the calculated energy and the quasi-bound state lies in the
middle of the stack, one gets a total transmission $T =1$.
In any other case the transmission is lower.

It remains to note,
that one should be careful with averaging over different probes.
The inherent fluctuation of the resistance can be reduced by averaging, but only in so far as the
resistivities of the different probes are the same.
This problem becomes apparent in the broad transition region to localization 
marked in \reffig{length_pic:3}.
Averaging of the
resistance gives the
impression of a non-linear dependence of the resistance on $L$.
\begin{figure}
\includegraphics[width=\picincolumnwidth]{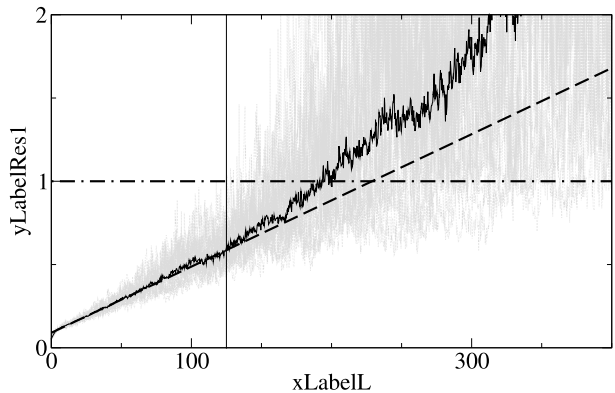}
\caption{\label{length_pic:4}%
Part of \reffig{length_pic:3} is shown with linear scaling of the resistance.
The start of the transition to localization is marked by the vertical line.
The linear extrapolation of \refeq{length_eq:1} is plotted as (-- --).
The fit was done for the inverse averaged conductance in the interval $(20 \dots 50)$ au.}
\end{figure}
This is best met by an average over the conductance (see \reffig{length_pic:4}), because
this corresponds to a parallel set-up of the probes. Compared to the averaging of the resistance,
samples with higher conductance are weighted stronger. 
One should note however, that a sample average for transmission above and below the threshold
$T=1$
does mean averaging over different `materials', i.e. metallic and insulating samples. 
As can be seen in \reffig{length_pic:4}, this leads to deviations from the linear resistance scaling
of \refeq{length_eq:1}.

This is the reason, why the resistivity calculations shown later rely on one larger sample, rather
then many small ones.
Even though the hybrid representation outlined above can be used for calculation of large stack
length, the main advantage is to increase the lateral length and thereby to reduce the difference
between samples in terms of the resistivity.
Just as for standard calculation of the resistivity by Kubo-Greenwood formula, one has to compute 
only one sample.
\subsection{\label{resisty}Calculation of the resistivity}
Due to the periodic boundary conditions in the lateral direction the conductance of
\refeq{res_eq:1} and \refeq{res_eq:2} depends on the $\kpar$-point in the resulting two
dimensional Brillouin zone.
Usually the calculations of the resistance is done in $\Gamma$-point approximation, that is only 
one
point in the middle of the two dimensional Brillouin zone is used to calculate 
the current through the sample.

\begin{figure}[b]
\includegraphics[width=\picincolumnwidth]{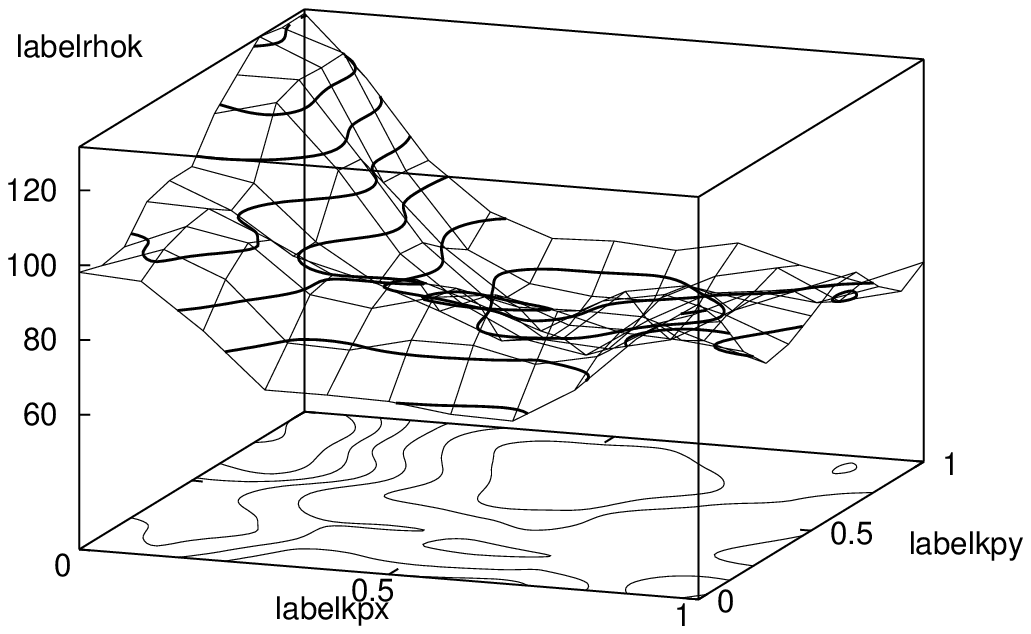}
\caption{\label{resisty_pic:1}%
Resistivity of liquid iron calculated for single $\kpar$-points of the two dimensional BZ based on 
\refeq{res_eq:1}.
The contour lines are plotted every $10$ \mOm~starting at $70$ \mOm.
The lateral length of the sample is $a=20$ au.}
\end{figure}

\begin{figure}
\includegraphics[width=\picincolumnwidth]{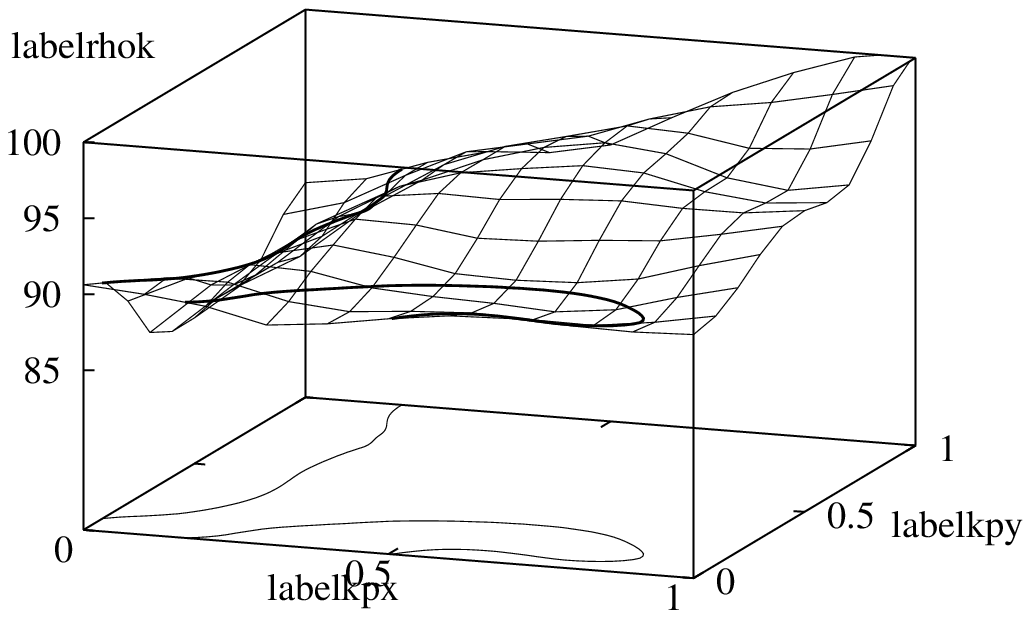}
\caption{\label{resisty_pic:2}%
Resistivity of liquid iron calculated for single $\kpar$-points as shown in \reffig{resisty_pic:1}
but with a lateral length $a=40$ au.
The contour line is plotted for $90$ \mOm}
\end{figure}

The error made by the single $\kpar$-point approximation is illustrated in \reffig{resisty_pic:1} for a
small lateral length and in \reffig{resisty_pic:2} for the doubled lateral length.
The differences in the resistivity decrease by increasing the lateral length $a$.
For small $a$ the resistivity values differ in the same order of magnitude as for different samples
in the $\Gamma$-point approximation.

In general the conductance has to be averaged over all $\kpar$-points\cite{Kahnt95}
\begin{equation}
\label{resisty_eq:1}
\bar{G} =\frac{A}{4\pi^2} \int\limits_\mathrm{BZ} \!\!\! d^2\kpar \,G(\kpar)
\end{equation}
This $\kpar$-dependence is in so far important, that
there are cases, when large lateral length $a$ are not attainable
and the accuracy is not sufficient.
An example is a small approximants of a quasi crystal,\cite{Solbrig00}
where the structure is well defined.
Apparently an averaging over samples is not possible.  
In such a case one has to use \refeq{resisty_eq:1}.

It should be noted however, that contrary to different samples,
the dependence on $\kpar$ in \reffig{resisty_pic:1} 
is far from random. 
Eventually the resistivity
is dominated by small regions with small resistivity.
Thus one has to calculate many different $\kpar$-points to get an accurate result, 
which makes this sampling more time consuming then a calculation of one large sample in 
$\Gamma$-point approximation.

Both examples shown here have the maximal value of the resistance near the far edges of the BZ.
Whereas the $\Gamma$-point gives a fair approximation of the averaged resistivity based on 
\refeq{resisty_eq:1}.
The $\kpar$-dependence is smoother for the larger lateral length. Hence a smaller number of sampling
points is sufficient to approximate the integral in \refeq{resisty_eq:1}.

For the calculations corresponding to \reffig{resisty_pic:1} the
resistivity is $83$ \mOm, compared to $92$ \mOm\ for the larger sample of \reffig{resisty_pic:2}.
Note however, that other samples with $a=20$ au showed up to $15$ \mOm\ higher resistivities.

To conclude this section we note, that for temperatures different from zero the conductance has to be
averaged over energies different from the Fermi energy.
Due to the phase shift of the scatterers and the propagation of the waves in the stack, the
conductance depends on the energy. 
Different energies have to be taken into account especially for higher temperatures.
This results in an average of \refeq{res_eq:1} weighted by the Fermi distribution.\cite{Datta92}
Whether calculations for different energies are necessary depends on the change of the current
transmission in an energy interval of order $\kb T$ around the Fermi energy.
For the liquid transition metals the conductance does not change much on the energy scale $\kb T$
and the average can be approximated by
\begin{equation}
\label{resisty_eq:2}
G (T) =\int_0^\infty\!\!\! dE \left(  - \frac{\partial f}{\partial E} \right) G(E)
\mathp
\end{equation}
As an example the temperature dependent resistivity and the resistivity coefficient is calculated 
for liquid sodium in section~\ref{weak}.
%
%
%
%
\section{\label{example}Application on different material classes}
\subsection{\label{transmet}Liquid 3d transition metals}
In the following section results of resistivity calculations are compared with standard methods,
i.e. Kubo-Greenwood and Ziman formula.
To ensure that variations in the results stem from the methods not from the input data, the same 
self-consistent LMTO program was used for the calculation of the scattering potential.\cite{Arnold95}
Because the super cells are usually much smaller then the stacks, the potential has to be 
averaged over all atoms of the same type.

The structures were generated by Reverse Monte Carlo method based on measured 
structure factors for the melting point.\cite{Waseda80}
The structure factor was matched closely by the RMC method aside from liquid titanium.

\begin{figure}[b]
\includegraphics[width=\picincolumnwidth]{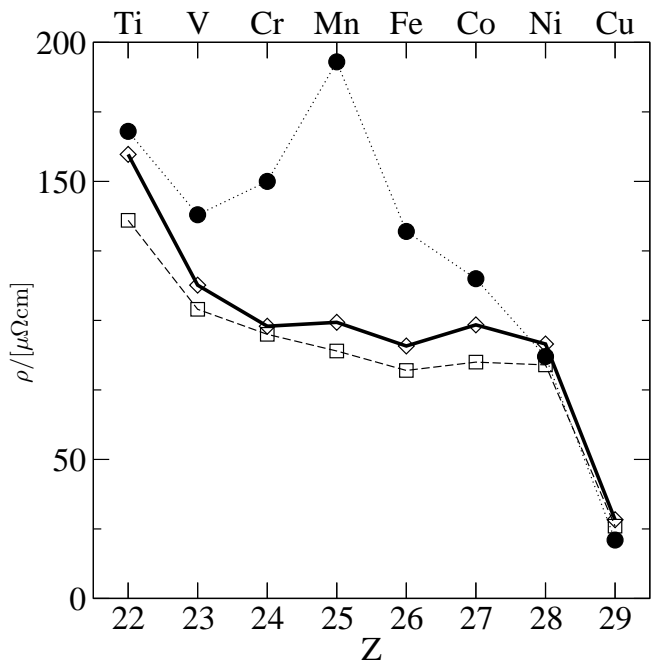}
\caption{\label{transmet_pic:1}%
Resistivity for liquid 3d transition metals and copper near the melting point. 
The calculations based on
\refeq{res_eq:1} ($\diamond$) are compared with Kubo-Greenwood results ($\square$) and measurements
($\bullet$).\cite{Arnold96}
Values from left to right in \mOm: 
$159.7;\, 112.7;\, 97.9;\, 99.3;\, 90.8;\, 98.4;\, 91.5;\, 28.3$.}
\end{figure}
In \reffig{transmet_pic:1} the resistivities of liquid 3d transition metals are compared with Kubo
calculations form \ocite{Arnold96} and experimental measurements from \ocite{Dyos92}.
The resistance calculations are done in $\Gamma$-point approximation by \refeq{res_eq:1}.
Because of the high melting temperatures for these metals, five different energies were used to
approximate \refeq{resisty_eq:2}.
Notable deviations of resistivity from the value at the Fermi energy were observed for liquid 
titanium and nickel.
This is caused by the strong change in the DOS around the Fermi energy for nearly filled or empty 
d band.

All stacks had a lateral length of 40 au and the linear regression was done between 10 au and 100
au.
In favor of a much faster calculation, a full projection on plane waves
was done only after every 80th layer.
The hybrid basis consisted of 9 angular momentum channels 
for each of the last 80 attached layers and about 320 plane waves.
Both values were tested by increasing the number of channels and by the accuracy of the
current conservation.

The resistivities calculated by linear regression of the multi-channel B\"uttiker formula 
\refeq{res_eq:1} are compared with Kubo-Greenwood resistivities in \reffig{transmet_pic:1}.
For completeness the experimental values are shown as well.
The differences between numerical and experimental results are discussed in \ocite{Arnold96}.
Essentially this is caused by the lack of spin dependence in the calculations.
Thus the differences to experiments are larger for half filled d band.

To demonstrate, that the agreement with Kubo-Greenwood resistivities is not limited to 
dominating d-scattering,
the resistivity of liquid copper is added in \reffig{transmet_pic:1}. 
Both numerical calculations are in good agreement with experimental values.\cite{Dyos92}

Agreement between the resistivity based on \refeq{length_eq:1} and the Kubo-Greenwood resistivity is
remarkably good for all liquid 3 d transition metals.
This shows that both methods are equally suited for resistivity calculations of strong scattering
disordered materials.
The only notable deviation is the up to 10 \mOm\ larger resistivity for the results by multi-channel
B\"uttiker formula.
Likewise there is no notable difference between the resistivities based on \refeq{res_eq:1} and
\refeq{res_eq:2}.
Both calculations use similar structure and scattering potentials, periodic boundary conditions in
lateral direction and are based on a linear response ansatz.
Likewise the incident current distribution has no effect on the resistivity.
The only two differences, which may have an effect on the resistivity, are the different 
lateral dimension and
the attenuation used for the Kubo-Greenwood calculation in \ocite{Arnold96}.

A similar difference for the resistivity occurred for the smaller samples used in section \ref{length} 
in case of liquid iron.
It should be noted, that the size
of the supercells for Kubo-Greenwood calculations was about the same as the lateral length 
$a\approx 20$ au 
of the smaller samples.

These differences in the resistivities resemble the quantum mechanical 
corrections due to multiple scattering in three dimensional finite systems,\cite{Lee85} 
\begin{equation}
\label{resisty_eq:3}
\sigma = \sigma_0 - \frac{e^2}{\hbar \pi^3} \left( \frac{1}{l_\mathrm{el}} - \frac{1}{a} \right)
\mathk
\end{equation}
where $a$ is the size of a cubic sample and $l_\mathrm{el}$ is the elastic mean free path.

For the example of liquid iron the Joffe-Regel limit is reached implying 
$l_\mathrm{el} \approx 5$ au.\cite{Arnold96}
This gives a weak localization correction for the resistivity of about $8$ \mOm,
which is of the same order compared to the 
$7$ \mOm\
higher resistivity for liquid iron in \reffig{transmet_pic:1}.
The increase in resistivity for the doubled lateral length $a$ is 
$1$ \mOm\ according to \refeq{resisty_eq:3} and is therefore much too small to account for 
the differences in \reffig{transmet_pic:1}.

Hence the main reason for the discrepancies seems to be the attenuation of the scattered waves
in \ocite{Arnold96},
which may have an effect on the weak localization correction.
One should note however, that the discussed discrepancies are in the range of the error margin
especially for small lateral length.
\subsection{\label{weak}Application for weak scattering}
Previous resistivity calculations based on \refeq{res_eq:1} were limited to fairly strong scattering,
because of the limited sample size.\cite{Kahnt95}
This constraint is relaxed by using the hybrid representation. 

\begin{figure}
\includegraphics[width=\picincolumnwidth]{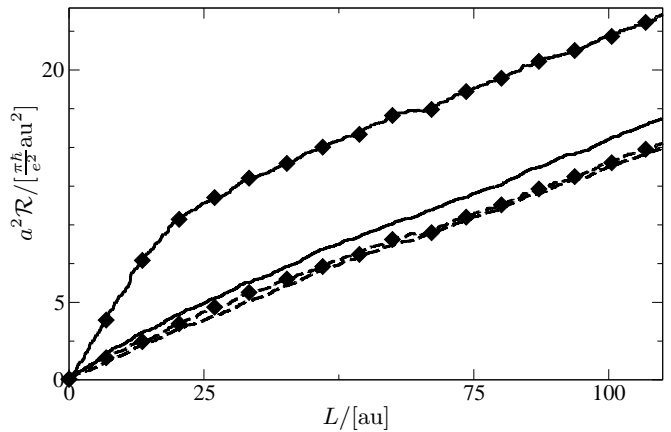}
\caption{\label{weak_pic:1}%
Resistance over sample length $L$ for two different lateral length ($a = 50, 70$ au).
The sample with the smaller lateral length of $50$ au is marked by $\blacklozenge$.
Two different resistance formulas are used, which correspond to ideal leads \refeq{res_eq:1} and
adaptive leads \refeq{res_eq:2} leads respectively.
The latter resistance values, i.e. \refeq{res_eq:2}, are plotted as broken lines for 
both lateral lengths.
Calculations were done in $\Gamma$-point approximation.
}
\end{figure}

Different resistance calculations for liquid sodium near the melting point are shown 
in \reffig{weak_pic:1}.
The transient region is much larger then for strong scattering systems 
(comp. \reffig{length_pic:1}).
The lateral length of 70 au was chosen according to the initial transient region for the
resistivity calculations.
The results was tested by samples with $a=50, 100$ au. 

Whereas for strong scattering there are only small differences between the resistance 
formula \refeq{res_eq:1} using 
ideal leads, 
and \refeq{res_eq:2}, 
the deviations become practically important for weak scattering.
For adaptive leads, i.e. \refeq{res_eq:2}, the transient region is smaller, because the incident
current distribution is already adapted to the scattering of the sample.

This adaption has to take place inside the sample for \refeq{res_eq:1}.
Due to the weak scattering,
long samples are needed to reach a current distribution independent from the 
even distribution of the incident currents.

The slope of \refeq{res_eq:1} in the transient regime
is larger then in the metallic regime. For instance a cubic cell with length $a=30$ au would give a
resistivity of about $45$ \mOm.

Although not
comparable, Kubo-Greenwood calculations with similar small cells give far to high resistivities 
near the melting point,\cite{Silvestrelli97}
whereas calculations based on Ziman's formula give good results near the melting 
point.\cite{Sinha89,Devlin72}

Increasing the lateral length increases the conductance per cross section based on \refeq{res_eq:1}, 
whereas values for \refeq{res_eq:2} are nearly independent of the lateral length even
in the transient region (see \reffig{weak_pic:1}).
Both equations have similar transient regions for a lateral length of $a=70$ au. 
Kubo-Greenwood calculation
with such a cell dimension should give similar results for the 
resistivity.

All the shown resistivities are calculated using \refeq{res_eq:2}, because of 
the smaller dependence on the lateral length $a$ (see \reffig{weak_pic:1}). 

The resistivity calculated for single $\kpar$-points differ by about $2$ \mOm\ for $a=70$ au
($3$ \mOm\ for $a=50$ au).
Sampling was done with 16 different $\kpar$-points to approximate \refeq{resisty_eq:1}.
The difference to samples with $a=50$ au is reduced below $0.4$ \mOm.
Such a high accuracy is necessary for the calculation of the resistivity coefficient near the
melting point
\begin{equation}
\label{weak_eq:1}
\alpha \approx \frac{1}{\rho(T_m)}\frac{\rho (T) - \rho(T_m)}{T - T_m}
\mathp
\end{equation} 

\begin{table}
\caption{\label{weak_tab:1}%
Calculated values of the resistivity near the melting point for liquid sodium are compared to
Kubo-Greenwood formula (KB) and Topp-Hopfield (TH) pseudopotential,\cite{Silvestrelli97}
Ziman formula with Shaw's pseudopotential (TW),\cite{Devlin72}
and measurements.\cite{Freedman61}
The present calculations are based on the same self-consistent calculated LMTO potential.
The extended Ziman formula\cite{Evans71} is compared with 
linear regression results for \refeq{length_eq:1} based on \refeq{res_eq:2} (Lenk).
The higher angular momentum scattering was disregarded for the last two calculations (s). 
}
\begin{ruledtabular}
\begin{tabular}{ldccc}
T (K) & \multicolumn{1}{c}{$\rho$ (\mOm)} & method & potential 
\\\hline
400 & \sim 25 & KG & TH 
\\
373 & 8.6 & Ziman & TW 
\\\hline
378 & 9.5 & ext. Ziman & LMTO 
\\
378 & 9.2 & Lenk & LMTO 
\\\hline
378 & 10.1 & ext. Ziman & LMTO (s) 
\\
378 & 9.1 & Lenk & LMTO (s) 
\\\hline
373 & 9.44 & \multicolumn{2}{c}{Exp.} 
\end{tabular}
\end{ruledtabular}
\end{table}
It can be seen from table \ref{weak_tab:1}, that calculations based on \refeq{res_eq:2} are
close to the resistivities based on Ziman formula.\cite{Devlin72,Ashcroft66,Sinha89} 
To underline this, values for the resistivity are included using the extended
Ziman formula of Evans.\cite{Evans71}
Because the same phase shift and structure factor is used, the deviations are caused by multiple
scattering.

As for strong scattering the sample structures were determined by RMC method based on 
measured structure factors.\cite{Waseda80}
The phase shifts (up to $l=2$) were calculated self-consistently by a LMTO method.\cite{Arnold95}
Half the minimal distance between atoms was chosen for the muffin tin radius.

Naturally s-scattering is dominating the resistivity for liquid sodium, but nevertheless the
inclusion of p- and d-scattering yields to a notably reduction of the resistivity.
This effect is small near the melting point and is nearly canceled due to multiple scattering
(see \reftab{weak_tab:1}),
but becomes more important for higher temperatures (comp. \reftab{weak_tab:2}).

The calculated resistivities based on the extended Ziman formula as well as multiple scattering 
are in good agreement compared to former calculations based on the Ziman formula,\cite{Devlin72}
and in even better agreement
to the measured resistivities near the melting point.\cite{Freedman61,Ashcroft66}
But so far the Ziman formula failed to describe the temperature dependence.\cite{Sinha89} 

\begin{table}
\caption{\label{weak_tab:2}%
Values of the resistivity for higher temperatures compared to measured resistivities from
\ocite{Freedman61}. Abbreviations are the same as for \reftab{weak_tab:1}.}
\begin{ruledtabular}
\begin{tabular}{ldccc}
T (K) & \multicolumn{1}{c}{$\rho$ (\mOm)} & method & potential 
\\\hline
473 & 11.5 & ext. Ziman & LMTO 
\\
473 & 10.5 & Lenk & LMTO 
\\\hline
473 & 13.95 & ext. Ziman & LMTO (s) 
\\
473 & 11.8 & Lenk & LMTO (s) 
\\\hline
823 & 11.7 & ext. Ziman & LMTO 
\\
823 & 11.4 & Lenk & LMTO 
\\\hline
423 & 11.1 & \multicolumn{2}{c}{Exp.} 
\\
473 & 12.9 & \multicolumn{2}{c}{Exp.}
\\
823 & 28.56 & \multicolumn{2}{c}{Exp.} 
\end{tabular}
\end{ruledtabular}
\end{table}
As can be seen from \reftab{weak_tab:2}, the deviations from measured resistivities increase with
increasing temperature, because the calculated resistivities remain comparatively unchanged.
Nevertheless, there is the good agreement between the multiple scattering results based on 
\refeq{length_eq:1}
and \refk{res_eq:2} and the extended Ziman formula.

The resistivity coefficients are summarized in \reftab{weak_tab:3}.
The resistivity coefficients are lower compared to the extended Ziman
formula, if multiple scattering is included. This is the same behavior as for the resistivity.
Likewise, the resistivity coefficients are increased, if higher angular momentum channels are
excluded.

The results are comparable to the value based on measurements for the reduced number of channels.
Possible reasons are relative inaccurate phase shifts for higher angular momentum channels, 
because these phase shifts are quite small.
A hint is the strong dependence of the calculated phase shifts for higher angular
momentum channels on the muffin tin radius.
\begin{table}
\caption{\label{weak_tab:3}%
Values of the resistivity coefficient near the melting point.}
\begin{ruledtabular}
\begin{tabular}{ldddddd}
& \multicolumn{1}{c}{Exp.} & \multicolumn{1}{c}{Ziman} &  \multicolumn{2}{c}{ext. Ziman}  
& \multicolumn{2}{c}{Lenk} \\\hline
$\alpha \cdot K/10^{3}  $ & 3.5 & 2.9 & 2.2 & (4.0) & 1.5 & (3.1)\\
\end{tabular}
\end{ruledtabular}
\end{table}

Anyway the agreement may be accidental, because both the extended
Ziman formula and the presented model fail to describe the temperature dependence of the
resistivity over a larger temperature range (see \reftab{weak_tab:2}).
\subsection{\label{insul}Resistance for insulating materials}
%
So far it was shown, that there exists a region, where the resistance scales linear with 
the sample length (apart from UCF).
For large sample length, there is a transition to localization due to disorder. 
This transition is caused by the lateral confinement. Thus the localization length
$l_\mathrm{loc} \approx N l_\mathrm{el}$ depends on the number of propagating modes $N$
and the elastic mean free path $l_\mathrm{el}$.\cite{Beenakker97}

\begin{figure}
\includegraphics[width=\picincolumnwidth]{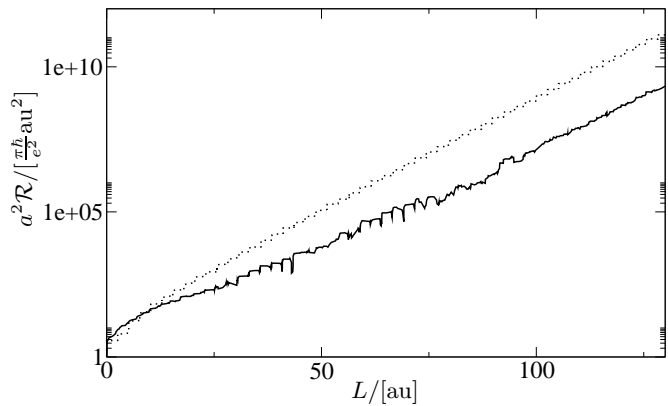}
\caption{\label{insul_pic:1}%
Resistance of amorphous (---) and crystalline silicon samples ($\dots$) in $\Gamma$-point
approximation in the middle of the gap (see \reffig{insul_pic:2}).
The lateral length of the samples is about $20$ au for a-Si
and c-Si. An additional sample with $a\approx 10$ au was calculated for c-Si, but it is
indistinguishable from the sample with $a\approx 20$ au in this representation.}
\end{figure}
In comparison the resistance is shown for samples of crystalline and amorphous silicon
in \reffig{insul_pic:1}.
The resistance increases exponentially in both cases, 
in contrast to metallic materials (see \reffig{length_pic:3}), 
where an exponential increase is only observed for the sample average.

The exponential scaling of the resistance is independent from the lateral length of the sample. 
The localization length is characteristic for the insulating material of the sample 
in contrast to metallic samples, where the localization length increases with increasing cross
section.

\begin{figure}[b]
\includegraphics[width=\picincolumnwidth]{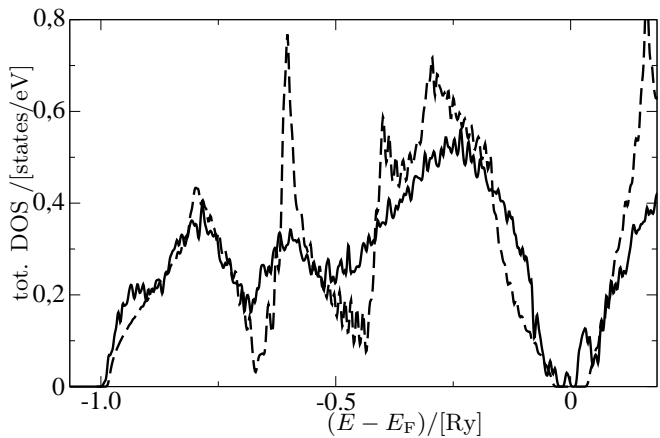}
\caption{\label{insul_pic:2}%
Electronic density of states for a-Si (---) and c-Si \mbox{(-- --)} calculated with a self-consistent 
LMTO method.}
\end{figure}
\begin{figure}
\includegraphics[width=\picincolumnwidth]{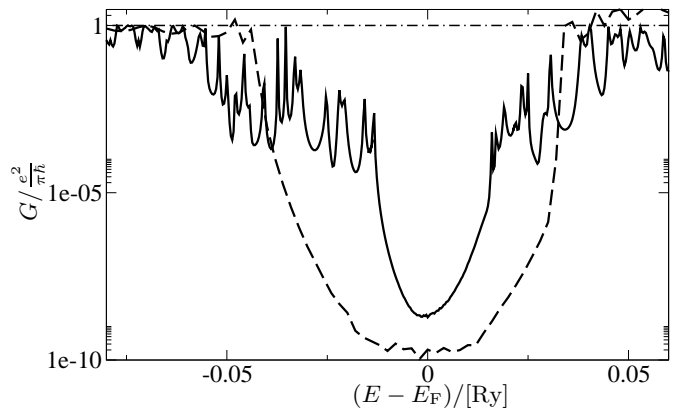}
\caption{\label{insul_pic:3}%
Energy-dependent conductance of the a-Si (---) and c-Si (-- --) samples shown in 
\reffig{insul_pic:1}. The length $L$ is about $160$ au for both samples.}
\end{figure}

Modelled structures show usually a distinctive metallic regime, because the energy of the electrons
is larger then the scattering potentials.
Therefore the exponentially decreasing current transmission has to result from a quite effective
destructive interference of the scattering waves.
Essentially these interferences result in a reduced density of states (DOS) near the Fermi energy,
which is connected to the diagonal parts of the inverse matrix in \refeq{scat_eq:3}.\cite{Arnold95}
The resulting gaps for crystalline and amorphous silicon are shown in \reffig{insul_pic:2}.
Note however, that there is no one to one correspondence between electronic defects in 
\reffig{insul_pic:2} and resonances of the transmission in \reffig{insul_pic:3}, because 
of the different employed boundary conditions and calculation methods.

Both DOS calculations in \reffig{insul_pic:2} were done with a standard self-consistent 
LMTO-method in atomic sphere approximation.
Only s- and p-electrons were used for the amorphous silicon calculation, because the cell included
512 silicon atoms (compared to 8 for the c-Si). 
Vacancies had to be included in the calculations to get better results for the DOS, 
since both silicon modifications are open structures.
About half the number of silicon
atoms was sufficient to get a reasonably filled structures.\footnote{An equal number of silicon
atoms and vacancies were used for c-Si calculations.}

The amorphous silicon networks were generated by a RMC method with subsequent relaxation by a MD
method. The first step was introduced to ensure an amorphous network structure with four fold
coordination. 
The network configuration was fitted to the radial pair distribution based on 
measured structure factor.\cite{Kugler93}
The latter step was necessary to get the gap at the Fermi energy and is based on multiple
scattering.\cite{Arnold98}

The combination of both methods for generating amorphous silicon structures will be described elsewhere.
The DOS is comparable to ab initio results.\cite{Car88}
The stability of the final structures was tested by an ab initio method (ABINIT) for a smaller sample
with 64 silicon atoms.\cite{Gonze02}
%
%
The energy dependent transmission through such an amorphous silicon network resembles the 
transmission through the crystalline counterpart at least in the middle of the gap 
(see \reffig{insul_pic:1} and \refk{insul_pic:3}).
Electronic defects in amorphous silicon 
are visible in the conductance especially for the tail region of the DOS (see \reffig{insul_pic:3}).

This part in the energy dependent conductance plot increases for larger sample length.
A conductance plot over the sample length in this energy region corresponds to the localized part 
of \reffig{length_pic:3}, where
the conductance is lower then $\Gfak$.
Only where the total transmission is lower then one, the energy dependent conductance can be
represented by a one dimensional model of coherent scattering.
%
%
%
%
%
%
\section{Conclusions}
Starting from scattered-wave calculations of mesoscopic quasi-one dimensional systems, 
the length dependence of the resistance can be divided in three different regimes.
For short length, the calculated resistance depends on the employed resistance formula.
For larger length the increment of the resistance becomes independent of the incident currents
and is characteristic for the material.

This second metallic regime can be used to calculate the resistivity for a wide range of disordered 
materials, because the resistance shows a nearly classical dependence on the sample size.
It was shown, that the results are in accordance with standard methods like
Kubo-Greenwood and Ziman formula.
Likewise there was a reasonable agreement with measurements.
A large number of atoms can be calculated with the presented hybrid
representation of the wave field to get a high accuracy for the calculated resistivity.
The precision is further increased by sampling over different
$\kpar$-points. 

The resistance shows large fluctuations for large sample length
due to disorder induced localization of the scattered waves. 
The sample averaged resistance increases exponentially.
The total transmission through the stack is lower then one. The length dependence resembles a
disordered one dimensional system. 

For another coherence effect, the Universal Conduction fluctuations (UCF), it is shown, that
the theoretical predicted value $2/15$ for a quasi-one dimensional system is met quantitatively.
This can be used as a test, how well defined the material properties are.

For the example of strong scattering liquid transition metals a somewhat larger resistivity 
was observed, compared to Kubo-Greenwood calculations.
This was interpreted by weak localization corrections known from scaling in
three dimensional systems.
This underlines  the similarities between the scattering in (finite) three dimensional systems and
in the metallic regime of a quasi-one dimensional system.

As a second example the resistivity of weak scattering liquid sodium near the melting point 
was calculated and compared to
extended Ziman formula using the same structural and scattering properties.
The agreement was good, but multiple scattering corrections seem to be important 
especially for the resistivity coefficient.
Both the resistivity and the resistivity coefficient are lowered by multiple scattering.

Whereas the calculated resistivity near the melting point is close to measured values and even the 
resistivity coefficient may be explained,
measured resistivities for higher temperatures are not matched by the present calculations.

As a final application of the method, the transmission through insulating crystalline and amorphous
silicon was calculated.
The atomic models show an exponential decrease of the transmission like in case of a
simple one dimensional potential barrier.

The energy dependent transmission through the barriers was compared.
Differences between amorphous and crystalline samples occur at the band edges, where the 
disorder leads to quasi-bound states.
This region corresponds to the localized regime observed for length dependent resistance of
metallic samples.
%
%
%
%
%
\appendix
\section{Propagators}
\label{app_prop}
\subsection{angular momentum basis}
The propagation of the waves in angular momentum representation between muffin tins in different
layers is described by the interplanar structure constants 
\begin{equation}
\label{prop_anh_eq1}
G_{LL'}^{ji}  =  \sum_{\bar{L}} C(L|\bar{L}|L') G_{\bar{L} 0}^{ji}
= \sum_{\bar{L}} C(L'|\bar{L}|L) G_{0 \bar{L}}^{ji}
\mathk
\end{equation}
with
\begin{equation}
 G_{\bar{L} 0}^{ji} = \sum_{\Vec{r}_s^{(i)}} i k
h_l^{(1)} \left( \Vec{r}_j - \Vec{r}_s^{(i)} \right)
e^{i \kpar (\Vec{r}_s^{(i)} - \Vec{r}_i)}
\mathp
\end{equation}
The propagation between muffin tins inside a layer is described by
\begin{equation}
\label{prop_anh_eq2}
G'_{LL'} = \sum_{\bar{L}} C(L|\bar{L}|L') G'_{\bar{L} 0}
= \sum_{\bar{L}} C(L'|\bar{L}|L) G'_{0 \bar{L}} 
\end{equation}
with
\begin{equation}
G'_{\bar{L} 0} = \sum_{\Vec{r}_s^{(j)} \neq \Vec{r}_j} i k
h_l^{(1)} \left( \Vec{r}_j - \Vec{r}_s^{(j)} \right)
e^{i \kpar (\Vec{r}_s^{(j)} - \Vec{r}_j)}
\mathk
\end{equation}
and $C(L'|\bar{L}|L)$ the Gaunt coefficients.
A direct calculation of \refeq{prop_anh_eq1} and \refk{prop_anh_eq2}, 
as well as a computation in reciprocal space is not possible, therefore an
Ewald summation is used.\cite{Kambe672}
\subsection{Transformation between plane waves and angular momentum representation}

Transformation from incident plane waves in angular momentum basis gives the matrix elements
\begin{equation}
\label{prop_anh_eq3}
\left( \Mat{A} \right)_{jL\, \vta \alpha} =
\frac{1}{\sqrt{\kappa_\vta}}\sqrt{4 \pi} (-1)^m
  e^{i \kta (\Vec{r}_j - \Vec{r}_{-\alpha})} Y_{-L} (\evec{\kta})
\mathp
\end{equation}
The back transformation of the scattered waves is given by
\begin{equation}
\label{prop_anh_eq4}
(\Mat{B})_{\vta \alpha \, iL} = \frac{2 \pi}{k \sqrt{\kappa_\vta} A}
\sqrt{4 \pi } 
  e^{i \kta (\Vec{r}_\alpha - \Vec{r}_i)} Y_{L} (\evec{\kta}) 
\mathp
\end{equation}
The propagation of electrons without scattering is best described in plane wave representation
\begin{equation}
\label{prop_anh_eq5}
(P^{hom})_{\vta \alpha \, \vta' \alpha'} =
\delta_{\alpha \alpha'} \delta_{\vta \vta'}
e^{i \Vec{k}_\vta^\alpha (\Vec{r}_r - \Vec{r}_l)}
\mathk
\end{equation}
where $\Vec{r}_\pm = \Vec{r}_{r/l}$ are the reference points right and left of the stack.
The derivation can be found in \ocite{Kahnt93}.
\section{Multiple Scattering in hybrid representation}
\label{app_hyb}
The projection of one part of the stack on plane waves basis 
\begin{equation}
\label{hy_eq1}
\left| \PLi{1} \rra = \Mat{A}_1 \left| \Ppi{1} \rra \quad ,\qquad
\left| \Pps{1} \right\rangle = \Mat{B}_1 \left| \PLs{1} \rra
\end{equation}
is the starting point of hybrid method. The representation is distinguished by the index $p$ for
plane wave and $L$ for angular momentum basis.
For convenience the remaining part of the wave field is divided in the notation. 
The scattered wave for the initial step is given by
\begin{equation}
\label{multhy_eq1}
\left( \begin{array}{c}
\Pps{1} \\
\PLs{2} \\
\PLs{3}
\end{array} \right)
=
\left(
\begin{array}{lcr}
\T_{11}^{pp} & \T_{12}^{pL} & \T_{13}^{pL} \\
\T_{21}^{Lp} & \T_{22}^{LL} & \T_{23}^{LL} \\
\T_{31}^{Lp} & \T_{32}^{LL} & \T_{33}^{LL} \\
\end{array}
\right)
\left( \begin{array}{c}
\Ppi{1} \\
\PLi{2} \\
\PLi{3}
\end{array} \right)
\mathp
\end{equation}
The additional substacks are attached by increasing the angular momentum basis
\begin{widetext}
\begin{eqnarray}
\label{hy_eq2}
\left(
\begin{array}{c}
\Pps{1} \\
\PLs{2} \\
\PLs{3} \\
\PLs{4}
\end{array}
\right) 
& = &
\left(
\begin{array}{cccc}
\T_{11}^{pp} & \T_{12}^{pL} & \T_{13}^{pL} & \Mat{0} \\
\T_{21}^{Lp} & \T_{22}^{LL} & \T_{23}^{LL} & \Mat{0} \\
\T_{31}^{Lp} & \T_{32}^{LL} & \T_{33}^{LL} & \Mat{0} \\
\Mat{0} & \Mat{0} & \Mat{0} & \T_{44}^{LL}
\end{array}
\right)
\left(
\begin{array}{c}
\Ppi{1} \\
\PLi{2} \\
\PLi{3} \\
\PLi{4}
\end{array}
\right) + 
\\
&+& \left(
\begin{array}{cccc}
\Mat{0} & \Mat{0} & \Mat{0} & \T_{11}^{pp} \Mat{P}_{14}^{pL} + \T_{12}^{pL} \Mat{P}_{24}^{LL} +
\T_{13}^{pL} \Mat{P}_{34}^{LL}\\
\Mat{0} & \Mat{0} & \Mat{0} & \T_{21}^{Lp} \Mat{P}_{14}^{pL} + \T_{22}^{LL} \Mat{P}_{24}^{LL} +
\T_{23}^{LL} \Mat{P}_{34}^{LL}\\
\Mat{0} & \Mat{0} & \Mat{0} & \T_{31}^{Lp} \Mat{P}_{14}^{pL} + \T_{32}^{LL} \Mat{P}_{34}^{LL} +
\T_{33}^{LL} \Mat{P}_{34}^{LL}\\
\T_{44}^{LL} \Mat{P}_{41}^{Lp} & \T_{44}^{LL} \Mat{P}_{42}^{LL} & \T_{44}^{LL} \Mat{P}_{43}^{LL} & \Mat{0}
\end{array}
\right)
\left(
\begin{array}{c}
\Pps{1} \\
\PLs{2} \\
\PLs{3} \\
\PLs{4}
\end{array}
\right) \nonumber
\mathk
\end{eqnarray}
\end{widetext}
where $\T_{44}^{LL}$ describes the scattering by the isolated substack and 
\begin{eqnarray}
\label{hy_eq2b}
\Mat{P}_{14}^{pL} &=& \left[ \Mat{P}_{14}^{hom} \Mat{B}_4 \right]^{-} \mathk\quad
\Mat{P}_{41}^{Lp} = \left[ \Mat{A}_4 \Mat{P}_{41}^{hom} \right]^{+} \mathk\quad
\nonumber\\
\Mat{P}_{m4}^{LL} &=& \Mat{G}_{m4}^{ji} \mathk\quad
\Mat{P}_{4m}^{LL} = \Mat{G}_{4m}^{ji} \mathk\quad m=2,3
\mathk
\end{eqnarray}
are the propagators between the old main stack and the sub stack (s. appendix~\ref{app_prop}).
The index $\alpha = \pm$ indicates the direction of propagation.

In the third step \refeq{hy_eq2} is solved, which results in a hybrid-matrix $\Mat{R}$,
\begin{equation}
\left| \Psi^\mathrm{sc} \rra = \Mat{R} \left| \Psi^\mathrm{inc} \rra
\mathp
\end{equation}
In the last step,  parts of the resulting hybrid-matrix
are projected on the plane wave basis
\begin{widetext}
\begin{equation}
\label{hy_eq3}
\left( \begin{array}{c}
\Pps{1+2} \\
\PLs{3} \\
\PLs{4}
\end{array} \right)
=
\left(
\begin{array}{ccc}
\R_{11}^{pp}+\R_{12}^{pL}\Mat{A}_2+\Mat{B}_2(\R_{21}^{Lp}+\R_{22}^{LL}\Mat{A}_2 )
& \R_{13}^{pL}+\Mat{B}_2\R_{23}^{LL} & \R_{14}^{pL}+\Mat{B}_2\R_{24}^{LL}\\
\R_{31}^{Lp}+\R_{32}^{LL}\Mat{A}_2 & \R_{33}^{LL} & \R_{34}^{LL} \\
\R_{41}^{Lp}+\R_{42}^{LL}\Mat{A}_2 & \R_{43}^{LL} & \R_{44}^{LL} \\
\end{array}
\right)
\left( \begin{array}{c}
\Ppi{1+2} \\
\PLi{3} \\
\PLi{4}
\end{array} \right)
\mathk
\end{equation}
\end{widetext}
and the right reference point is moved to the right of the stack.
The hybrid representation of the increased stack has the same structure and 
dimension as the initial stack (see \refeq{multhy_eq1}).
Projection of the remaining parts on (propagating) plane waves gives \refeq{scat_eq:3}.

Repetitive applying of these four steps gives the wave field of the successive increased stack
without increase of the numerical cost per step.
The accuracy were tested by current conservation. Stacks with up to 10000 d-scatterers were
calculated.
The in general increasing error in the current conservation is not dominated by the above inversion
but by the Ewald summation for the structure constants of appendix~\ref{app_prop}.

\bibliography{paper}

\end{document}